\documentclass[twocolumn, twocolappendix]{aastex701}
\usepackage{amsmath}
\usepackage{changepage}
\usepackage{subcaption}
\usepackage{placeins}  % in preamble
\numberwithin{equation}{section}

\usepackage{hyperref}
\usepackage{natbib}

\begin{document}

% \title{Orbital evolution of Gigahertz Peaked Spectra in binary MSP using Parkes Radio Telescope}
\title{Evidence of Orbital Evolution of Gigahertz-Peaked Spectra in Binary Millisecond Pulsar}
% \title{Systematic Orbital-Phase Evolution of Gigahertz-Peaked Radio Spectra in Binary Millisecond Pulsars}

\author[0009-0001-9428-6235]{Rahul Sharan}
\affiliation{National Centre for Radio Astrophysics, NCRA-TIFR, Pune 411007, India}
\email[show]{rsharan.ncra@gmail.com, rsharan@ncra.tifr.res.in}

\author[0000-0002-6287-6900]{Bhaswati Bhattacharyya}
\affiliation{National Centre for Radio Astrophysics, NCRA-TIFR, Pune 411007, India}
\email{bhaswati@ncra.tifr.res.in}

\begin{abstract}
We present the first systematic investigation of the continuous orbital evolution of the  Gigahertz-Peaked Spectum turnover frequency ($\nu_{peak}$) in the binary millisecond pulsar, PSR J2144$-$5237. Using approximately 50 hours of densely sampled observations spanning the $\sim10$-day orbit with the Parkes Ultra-Wideband Low frequency receiver, we obtained continuous frequency coverage from 704 MHz to 4 GHz, enabling robust measurements of the spectral evolution throughout the binary orbit. We detect a clear and systematic modulation of the turnover frequency with orbital phase for both the $1^{st}$ pulse and $2^{nd}$ pulse components, with both emission components exhibiting remarkably similar evolutionary trends despite originating from distinct emission regions within the pulsar magnetosphere. The correlated evolution of the two pulse components strongly suggests that the observed spectral variability is governed by propagation through the intra-binary environment. The large orbital modulation of turnover frequency can be explained by cyclotron resonance absorption in magnetized plasma associated with the companion, while alternative absorption mechanisms are unable to reproduce the observed behaviour. Modelling the orbital evolution of the turnover frequency within the cyclotron absorption framework, can constrain the electron density distribution in the pulsar wind, estimate the magnetic field strength of the companion, and place limits on the geometry of the absorbing plasma and the binary system. This study establish orbital evolution of the Gigahertz Peaked Spectum as a powerful new diagnostic of the magnetized intra-binary medium and provide a hitherto unexplored probe of pulsar wind-companion interactions in eclipsing millisecond pulsar binaries.
\end{abstract}


\section{Introduction} \label{sec:Intro}
The radio spectra of pulsars provide important clues to the physical processes governing particle acceleration and coherent radio emission within their magnetospheres. Consequently, the spectral properties of pulsars have been extensively investigated over the past several decades to understand both the underlying emission mechanisms and the influence of the surrounding environment on the observed radio emission \citep{Sieber_1973, Maron_et_al_2000, Basu_et_al_2022}. The observed flux density is commonly described by a power-law relation, $S(\nu)\propto\nu^{\alpha}$, where $S(\nu)$ is the flux density at observing frequency $\nu$ and $\alpha$ is the spectral index. While most pulsars exhibit steep power-law spectra, deviations from this behaviour, particularly spectral turnovers, can provide valuable insights into the physical conditions within the pulsar environment and along the propagation path.

Traditional spectral studies generally assume that the propagation medium between the pulsar and the observer remains stable over the timescale of the observations. Under this assumption, the observed spectrum is considered to be predominantly intrinsic to the pulsar. However, growing observational evidence suggests that environmental effects can significantly modify the observed spectral shape \citep{Bruk_et_al_1978}. \cite{Kramer_et_al_1998, Dai_et_al_2015} and \cite{Qingzheng_et_al_2026} reported that millisecond pulsars (MSPs) may exhibit systematically different spectral indices compared to those of normal pulsars, motivating further investigations into the origin of their spectral diversity.

Several studies have since demonstrated that propagation effects can substantially influence pulsar radio spectra. For isolated pulsars, thermal free-free absorption by ionized filaments associated with supernova remnants or pulsar wind nebulae has been proposed to explain low-frequency spectral turnovers \citep{Lewandowski_2015}. In binary systems, however, the interaction between the pulsar wind and material lost from the companion introduces additional propagation effects that can significantly alter the observed spectrum. \cite{Kijak_et_al_2011a} carried out a systematic investigation of pulsars exhibiting Gigahertz Peaked Spectum (GPS) and found that many GPS pulsars are associated with unusually dense or complex environments, indicating that the spectral turnover is likely caused by external absorption rather than being an intrinsic characteristic of the radio emission mechanism. Building on this idea, \cite{Kijak_et_al_2011b} proposed that free-free absorption within the ionized stellar wind and cyclotron resonance absorption in magnetized plasma surrounding the companion can naturally reproduce the observed GHz-frequency spectral turnovers in binary pulsars. Such models suggests that the turnover frequency depends sensitively on the density, temperature, magnetic field strength, and path length through the absorbing medium. Subsequently, \cite{Rajwade_et_al_2015} presented broadband observations of several eclipsing spider MSPs and demonstrated that their radio spectra evolve with orbital phase, providing compelling observational evidence that the intra-binary plasma can dynamically modulate the observed radio emission. 
% Their work established that the absorbing medium is neither homogeneous nor static, but instead evolves throughout the orbit as the pulsar signal traverses different regions of the companion wind. Collectively, these studies established environmental absorption as the leading explanation for the GPS phenomenon, particularly in interacting binary MSPs.

Despite these important advances, the orbital evolution of the GPS turnover itself has remained largely unexplored. Previous investigations have primarily focused on identifying GPS pulsars or comparing spectra obtained at a limited number of orbital phases. Continuous tracking of the turnover frequency over an entire binary orbit has rarely been attempted, largely because most MSPs are intrinsically faint radio sources with steep spectra, making accurate wideband flux-density measurements challenging. Moreover, densely sampling an entire binary orbit with simultaneous broadband observations has only recently become feasible with instruments such as the Parkes Ultra-Wideband Low (UWL) receiver. Such observations provide a unique opportunity to probe the spatial distribution, density, and magnetization of the absorbing plasma through its influence on the evolving radio spectrum.

In \cite{Sharan_et_al_2026}, wideband observations of the spider MSP PSR J2144$-$5237 revealed compelling evidence for orbital evolution of its radio spectrum. The first column of Figure 3 of \cite{Sharan_et_al_2026} shows that the spectral turnover frequency evolves systematically throughout the orbit, shifting progressively from the higher-frequency end toward the lower-frequency end of the Parkes UWL band. This behaviour represents one of the first clear indications that the GPS characteristics of an MSP evolve continuously with orbital phase, suggesting that the observed spectrum is dynamically regulated by the changing properties of the intra-binary plasma rather than remaining temporally invariant.

Motivated by this context, the present paper investigates the orbital evolution of the GHz-peaked spectrum in PSR J2144$-$5237 in greater detail. By obtaining densely sampled observations across the binary orbit, this work examines whether the spectral turnover exhibits systematic orbital modulation, quantifies the evolution of the spectral parameters as a function of orbital phase, and explores the physical mechanisms responsible for the observed variability. Understanding these spectral changes provides a powerful diagnostic of the interaction between the pulsar wind and the companion outflow, while also offering new insight into the structure, density, and magnetic properties of the intra-binary plasma.

The remainder of this paper is organized as follows. Section \ref{sec:obs_n_data_analysis} describes the observations and data reduction procedures. Section \ref{sec:Results} presents the orbital evolution of the radio spectrum and the derived spectral parameters. Section \ref{sec:Discussion} discusses the implications of these results in the context of absorption and propagation mechanisms within the binary system, and Section \ref{sec:Future_work} outlines future directions for this work.

% \begin{figure}[!hbt]
%  \hspace*{-1.1cm}
%  \centering
%  \includegraphics[scale=0.9]{PNG/GPS_ex.png}
%  % \hspace*{1cm}
%  \caption{Evolution of the radio spectrum of the binary pulsar PSR J1302--6350 in the vicinity of the intra-binary medium, reproduced from \cite{Kijak_et_al_2011b}. The different coloured curves represent spectra measured at different orbital phases relative to periastron. For example, the green curve corresponds to the spectrum averaged over 113--186 days after periastron. \label{fig:GPS_ex}}
% \end{figure}
% MSP GPS : Larger MSP duty cycle implying more photons per unit time, to interact with extrinsic factors.

%  Layout of paper

\begin{figure*}[!hbt]
    \includegraphics[scale=0.6]{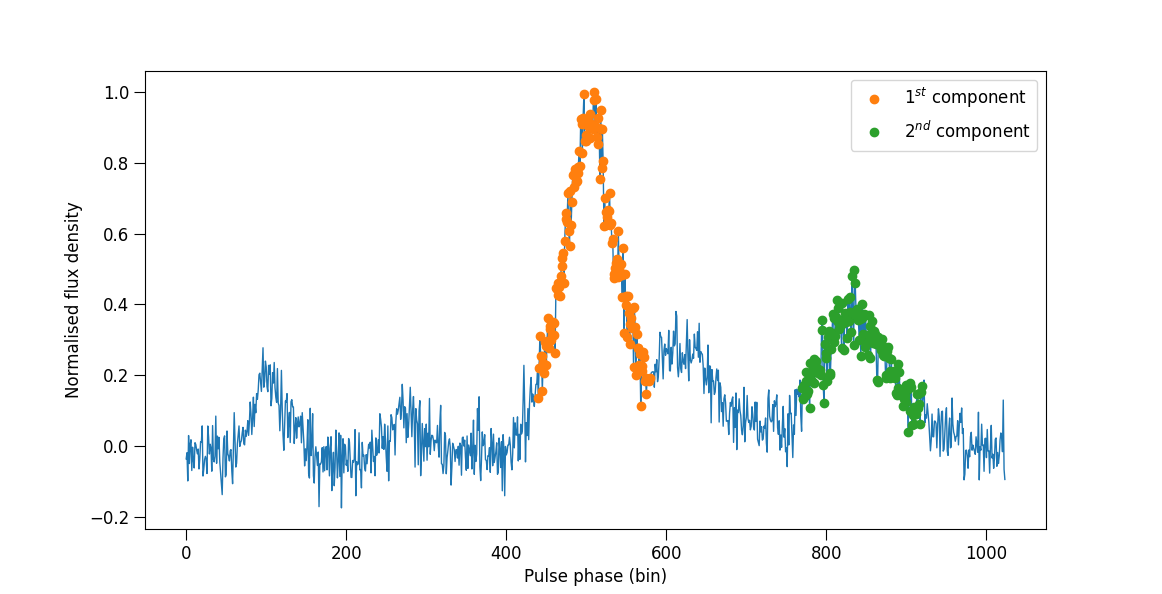}
\caption{The figure contains the average profile with $1^{st}$ and $2^{nd}$ components marked in the orange and green colored dots for the observing epoch 2023--06--22.}
\label{fig:Profile_J2144}
\end{figure*}

% \begin{figure*}[!hbt]
%     \begin{subfigure}[t]{0.2\textwidth}
%     % \hspace*{-0.9cm}
%       % \centering
%       \fbox{\includegraphics[scale=0.53]{PNG/GPS_evolution_J2144-5237.png}}
%       \caption{}
%       \label{fig:GPS_evolution_J2144}
%     \end{subfigure}
%     \begin{subfigure}[t]{1\textwidth}
%       % \centering
%       \hspace*{2.5cm}
%       \fbox{\includegraphics[scale=0.7]{PNG/Frequency_phase_plot_J2144.png}}
%       \caption{}
%       \label{fig:Frequency_phase_plot_J2144}
%     \end{subfigure}
% \caption{The left panel of the Figure shows dynamic spectra for the main component (shown in the dots), average profile and averaged spectral of J2144--5237 averaged for 20 minutes \cite{Sharan_et_al_2026}. The right panel shows the frequency versus spin phase (bin).}
% \end{figure*}

\section{Observational details and data analysis} \label{sec:obs_n_data_analysis}

%   Courser observation details
PSR J2144$-$5237 was observed for a total of approximately 50 hours using the Parkes Ultra-Wideband Low (UWL) \citep{Parkes_UWL} receiver, providing continuous frequency coverage from 704 to 4032 MHz. The pulsar is in a binary system with an orbital period of approximately 10 days, enabling extensive sampling of the orbit across multiple observing epochs. Details of the observations, including the observing dates, integration times, and orbital phase coverage, are summarized in Table~\ref{tab:nu_peak_comp_corr}.

%   PSRPYPE and PSRCHIVE calibration steps

All observations were acquired in fold mode with a sub-integration time of 30 s and a frequency resolution of 1 MHz. The data were processed using the \texttt{PSRPYPE}\footnote{https://github.com/vivekvenkris/psrpype} pipeline, which performs automated radio-frequency interference (RFI) excision using the \texttt{CLFD} algorithm, followed by full-Stokes polarization calibration through the \texttt{PSRCHIVE} software package. The calibrated archives were subsequently inspected to ensure data quality prior to spectral analysis.

%   Treatement of 20 minutes chunk to spectral : main & inter pulse spectra

To investigate the orbital evolution of the spectra, the calibrated data were divided into consecutive 20-minute integrations, corresponding to narrow orbital-phase intervals while maintaining sufficient signal-to-noise ratio for reliable spectral measurements. For each 20-minute integration, flux densities were measured across the UWL band separately for the two profile components:  the $1^{st}$ component (spin phase bin range : 440--580 in Figure \ref{fig:Profile_J2144}) and $2^{nd}$ component (spin phase bin range : 770--920 in Figure \ref{fig:Profile_J2144}), yielding a sequence of orbital phase-resolved spectra.

%   Fitting approach : broken power law, bayesian approach (Dynesty), resulting in posteriors 

The spectrum corresponding to each orbital phase segment was modelled using a Bayesian parameter estimation framework implemented with the nested sampling package \texttt{DYNESTY}. We adopted the broken power-law model described by Equation \ref{eq:Model} \citep[same as Equation 13 of ][]{Jankowski_et_al_2018}, which has been widely used to characterize pulsar spectra exhibiting GPS. The posterior distributions of the model parameters were sampled to estimate the spectral turnover frequency, $\nu_{peak}$ (represented as $\nu_0$ as model parameter), together with its associated uncertainties. 
\begin{equation}
    S(\nu) = 
    S\begin{cases}
    \nu^{\alpha_0}, & \nu \leq \nu_0 \\
    \nu_0^{\alpha_0 - \alpha_1}\nu^{\alpha_1}, & \nu > \nu_0
    \end{cases}
    \label{eq:Model}
\end{equation}
where parameter $\nu_0$ is the $\nu_{peak}$ (in GHz), $\alpha_0$ ($\alpha_1$) is the spectral index before (after) the turnovers, and $S$ is the overall scaling factor.

\begin{figure*}[h!]
    \begin{subfigure}[t]{0.5\textwidth}
      %\centering
      \includegraphics[trim={0 0 0 1.2cm}, clip, scale=0.45]{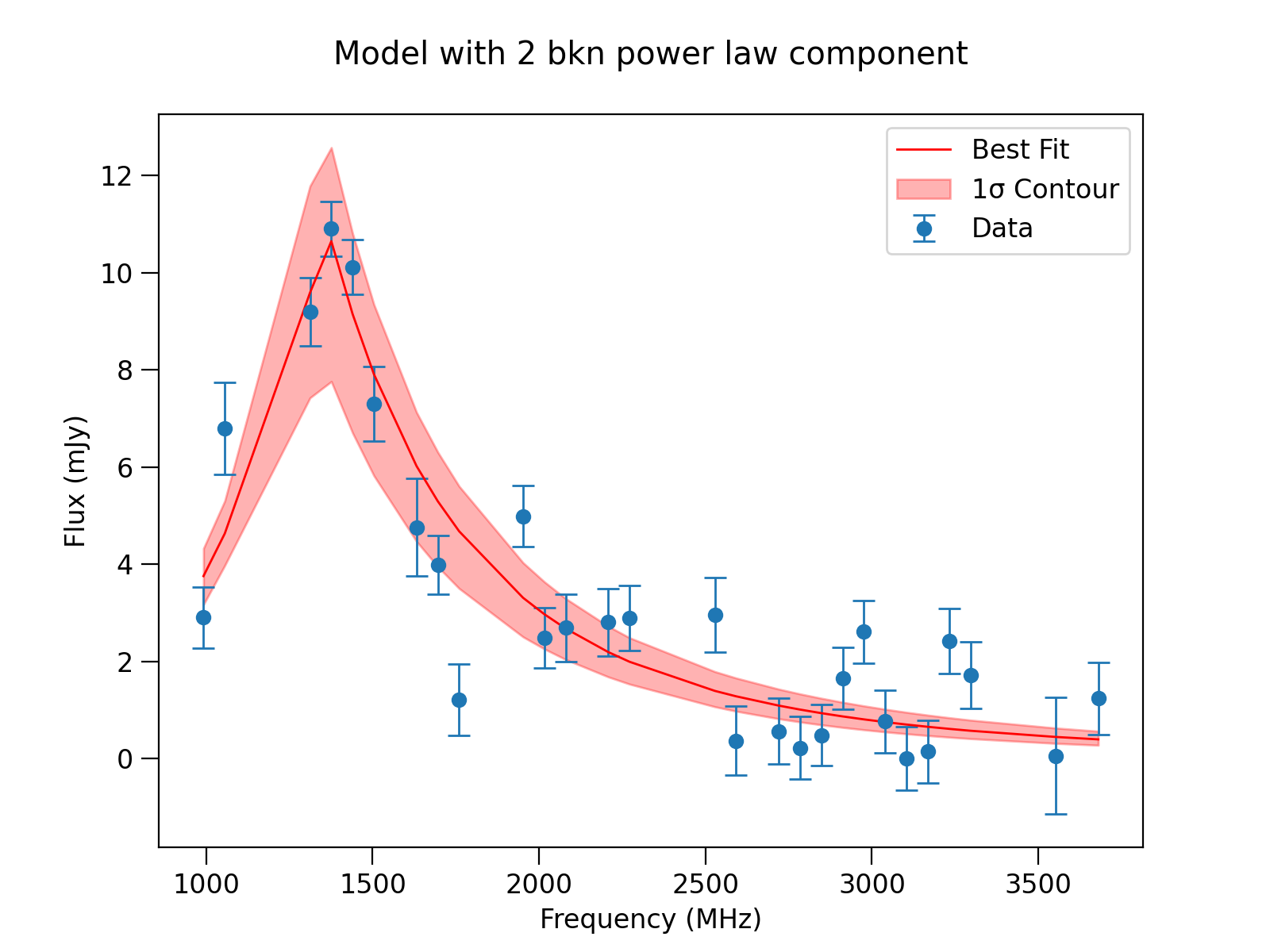}
      \caption{}
      \label{fig:Spectral_fit_plot}
    \end{subfigure}
    \begin{subfigure}[t]{0.5\textwidth}
      %\centering
      \hspace{-0.5cm}
      \includegraphics[scale=0.4]{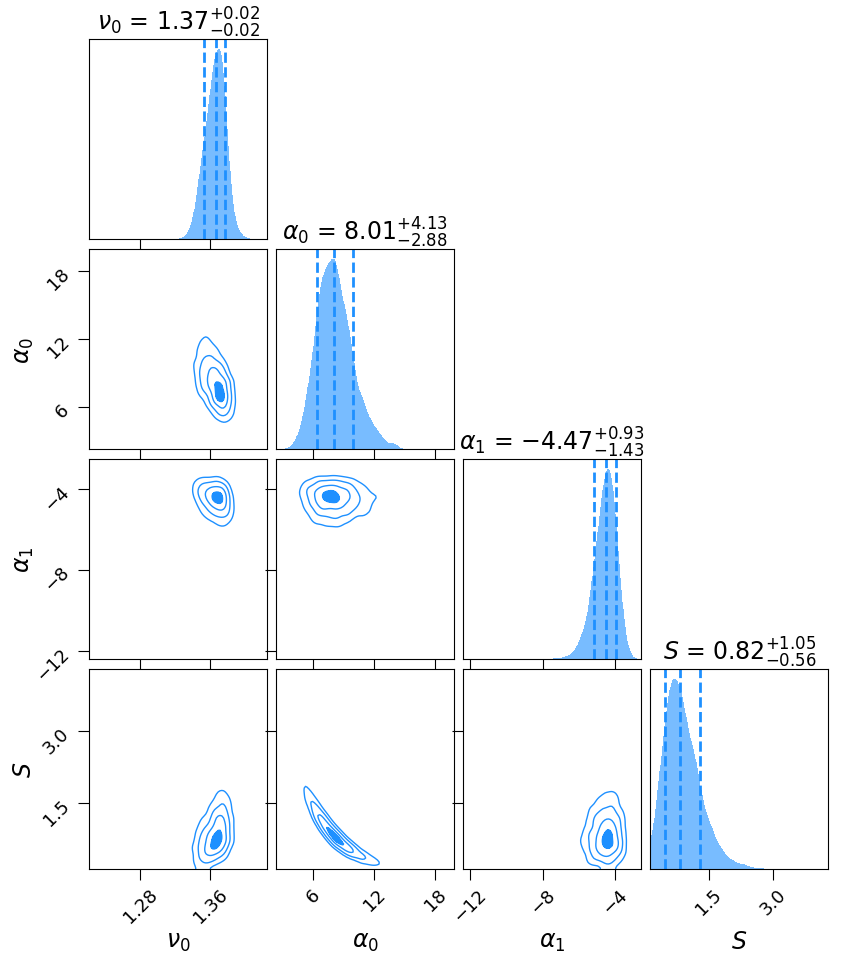}
      \caption{}
      \label{fig:Corner_plot}
    \end{subfigure}
\caption{Left panel contains a template plot which shows the GPS fit of a power law model (Equation \ref{eq:Model}) along with the data at the an orbital phase of 0.88 (observing epoch 2025--10--29). Right panel shows the corresponding corner plot.}
%, where parameter $\nu_0$ is the $\nu_{peak}$ (in GHz), $\alpha_0$ ($\alpha_1$) is the spectral index before (after) the turnovers, and $S$ is the overall scaling factor.}
\label{fig:GPS_eg_corner}
\end{figure*}

\begin{table*}[!h]
\centering
\begin{tabular}{|c|c|c|c|c|c|}
\hline
Epoch & observation duration (hrs) & orbital phase range &  Pearson correlation & P-value & Number of points \\
\hline
2025-10-29 & 4.91 & 0.875 -- 0.895 & 0.026 & 0.955 & 7 \\
2025-11-17 & 4.33 & 0.661 -- 0.678 & 0.976 & $2.61 \times 10^{-9}$ & 14 \\
2025-11-23 & 5.58 & 0.228 -- 0.251 & - & - & - \\
2025-12-05 & 5.17 & 0.362 -- 0.382 & -0.484 & 0.331 & 6 \\
2025-12-16 & 4.48 & 0.399 -- 0.418 & 0.759 & 0.006 & 11 \\
2026-01-30 & 3.86 & 0.644 -- 0.659 & 0.546 & 0.128 & 9 \\
2025-05-28 & 3.18 & 0.356 -- 0.369 & - & - & - \\
2023-08-26 & 8.84 & 0.738 -- 0.773& 0.697 & 0.0006 & 20 \\
2023-06-22 & 1.97 & 0.605 -- 0.613 & 0.702 & 0.078 & 7 \\
2023-05-29 & 8.52 & 0.350 -- 0.384 & 0.982 & $1.358 \times 10^{-5}$ & 8 \\
\hline
\end{tabular}
\caption{The table illustrates the pearson correlation of $\nu_{peak}$ for $1^{st}$ and $2^{nd}$ component. The reliable number of points (last column) shows reasonably better fitting results. }
\label{tab:nu_peak_comp_corr}
\end{table*}

%   Exceptions for broken power law, and edge cases due to requirement of complex modelling

% PSR J2144$-$5237 was observed for a total of approximately 50 hours with the Parkes Ultra-Wideband Low (UWL) receiver, providing continuous frequency coverage from 704 to 4032 MHz.

% The binary system has an orbital period of approximately 10 days, allowing the observations to sample a large fraction of the orbit across multiple epochs. The details of the observing epochs, including their durations and corresponding orbital phase coverage, are listed in Table~\ref{tab:nu_peak_comp_corr}. The observations was taken in the fold mode, with a 30s sub-integration time and 1 MHz channel width. Using \texttt{PSRPYPE}, we performed flagging (using \texttt{CLFD}) and full stokes observation calibration (using \texttt{PSRCHIVE}). 

% \begin{figure}[!hbt]
%  % \hspace*{-1.1cm}
%  \centering
%  \includegraphics[scale=0.35]{PNG/GPS_steps.png}
%  % \hspace*{1cm}
%  \caption{The figure shows the procedure followed for estimating multi epoch orbital phase resolved $\nu_{peak}$ for $1^{st}$ and $2^{nd}$ components.\label{fig:GPS_steps}}
% \end{figure}

\section{Results}
\label{sec:Results}

\begin{figure}[!hbt]
 % \hspace*{-1.1cm}
 \centering
 \includegraphics[trim={0 0 0 1.2cm}, clip, scale=0.4]{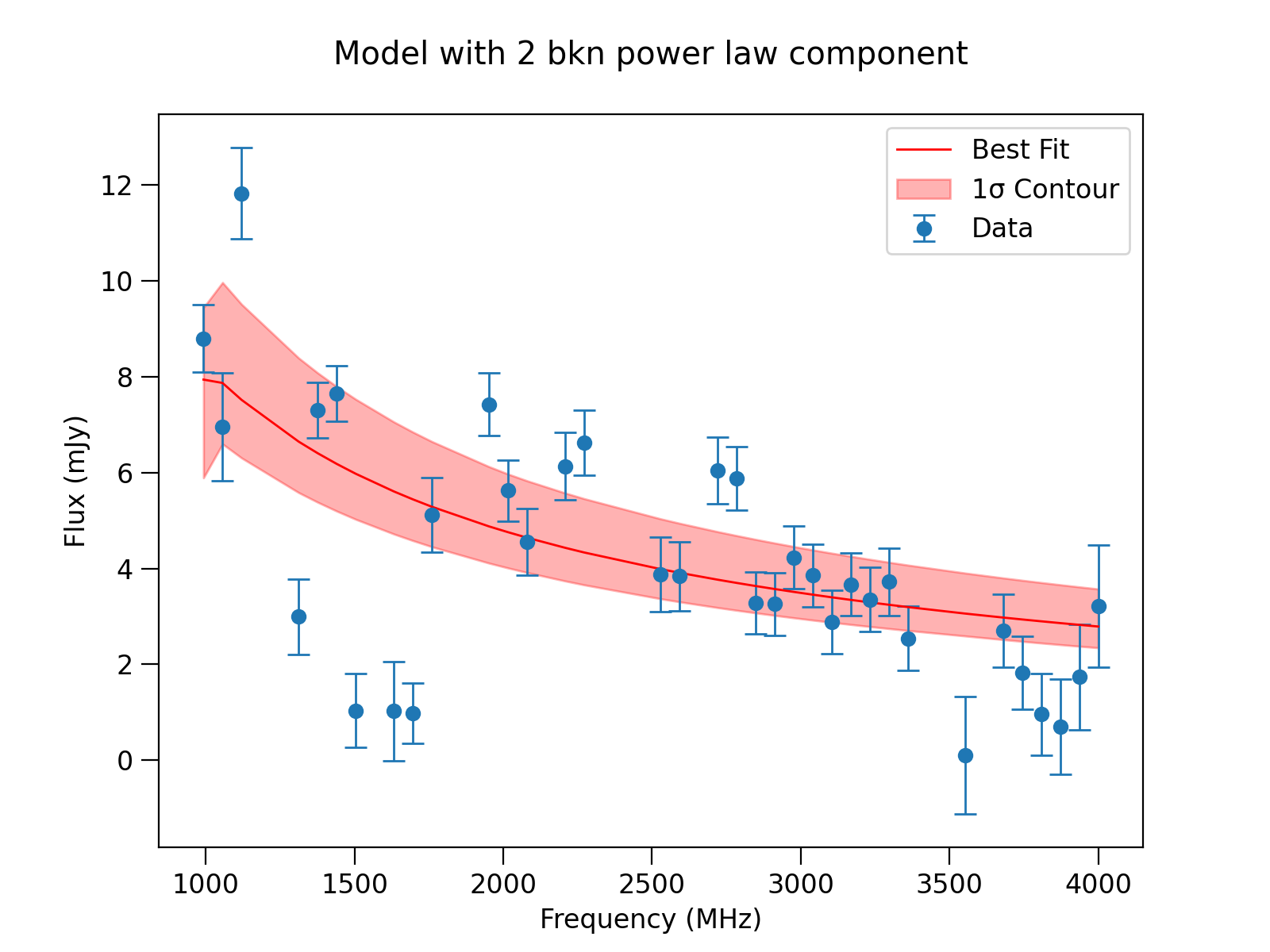}
 % \hspace*{1cm}
 \caption{The figure shows the model being too simplistic to capture the spectral features of a 20 minutes averaged data observed at orbital phase of 0.23 (observing epoch 2025--11--23).\label{fig:GPS_misfit}}
\end{figure}

An example of the spectral fit is shown in left panel of Figure~\ref{fig:GPS_eg_corner}, while the corresponding posterior distributions are presented in right panel of Figure~\ref{fig:GPS_eg_corner}. For a subset of orbital phases, the observed spectra exhibit more complex structure than can be adequately described by a simple broken power-law model, resulting in comparatively poorer fits (Figure \ref{fig:GPS_misfit}). This behaviour suggests that the spectral evolution is influenced by frequency-dependent propagation effects within the intra-binary plasma, indicating that physically motivated radiative transfer models incorporating absorption and emission processes may provide a more accurate description of the observed spectra. %Such modelling will be explored in future work.

%   Resulting explaining via the spectral data-model plots and corner plots

\begin{figure*}[!hbt]
    \centering
    % First row: Main component
    \begin{subfigure}[t]{0.48\textwidth}
        % \centering
        % \hspace{-2.0cm}
        \includegraphics[trim={0 0 0 1.2cm}, clip, scale=0.5]{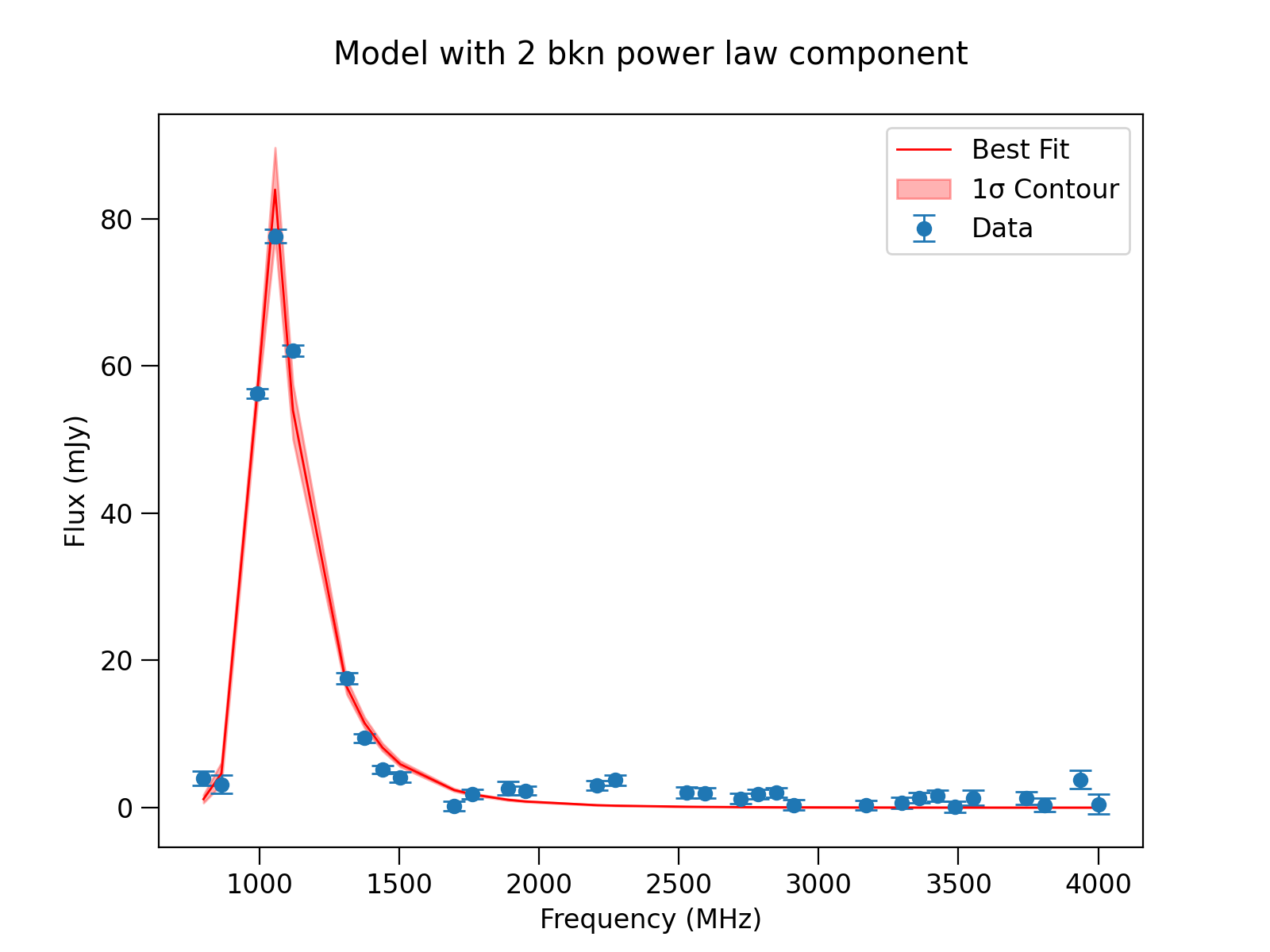}
        \caption{GPS spectral fit for the $1^{st}$ component.}
        \label{fig:Spectral_fit_plot_main}
    \end{subfigure}
    \hfill
    \begin{subfigure}[t]{0.48\textwidth}
        % \centering
        % \hspace{1.0cm}
        \includegraphics[scale=0.4]{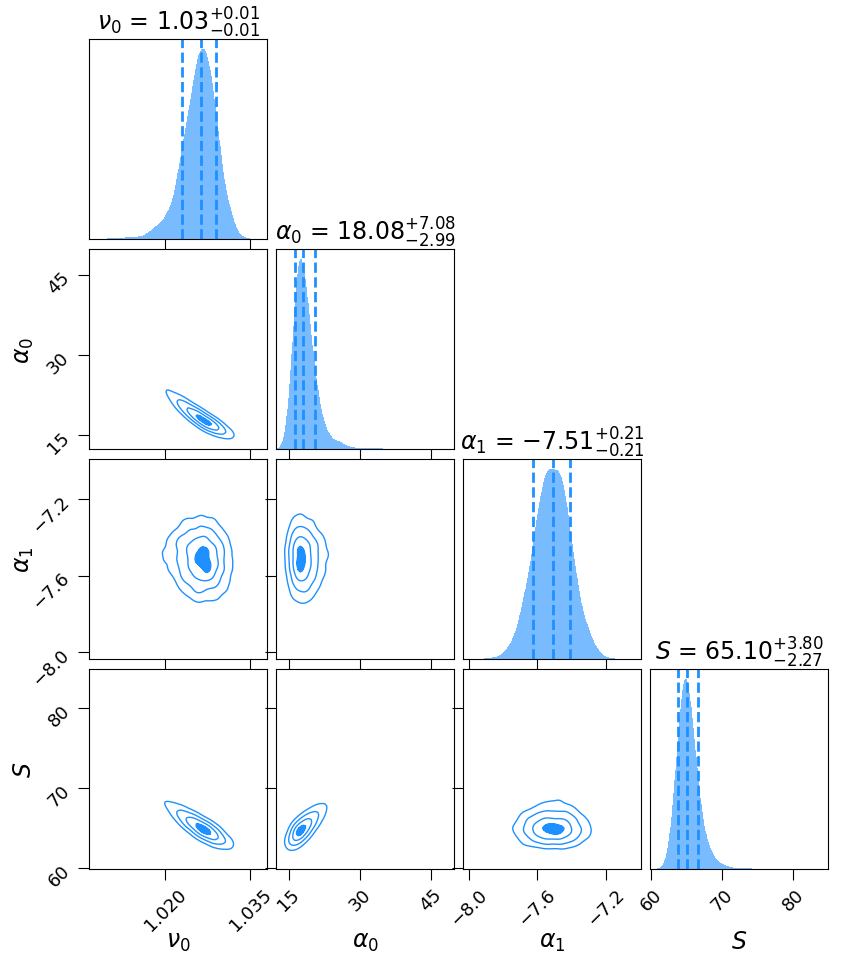}
        \caption{Corner plot for the $1^{st}$ component.}
        \label{fig:Corner_plot_main}
    \end{subfigure}

    \vspace{0.5cm}

    % Second row: second pulse component
    \begin{subfigure}[t]{0.48\textwidth}
        % \centering
        % \hspace{-2.0cm}
        \includegraphics[trim={0 0 0 1.2cm}, clip, scale=0.5]{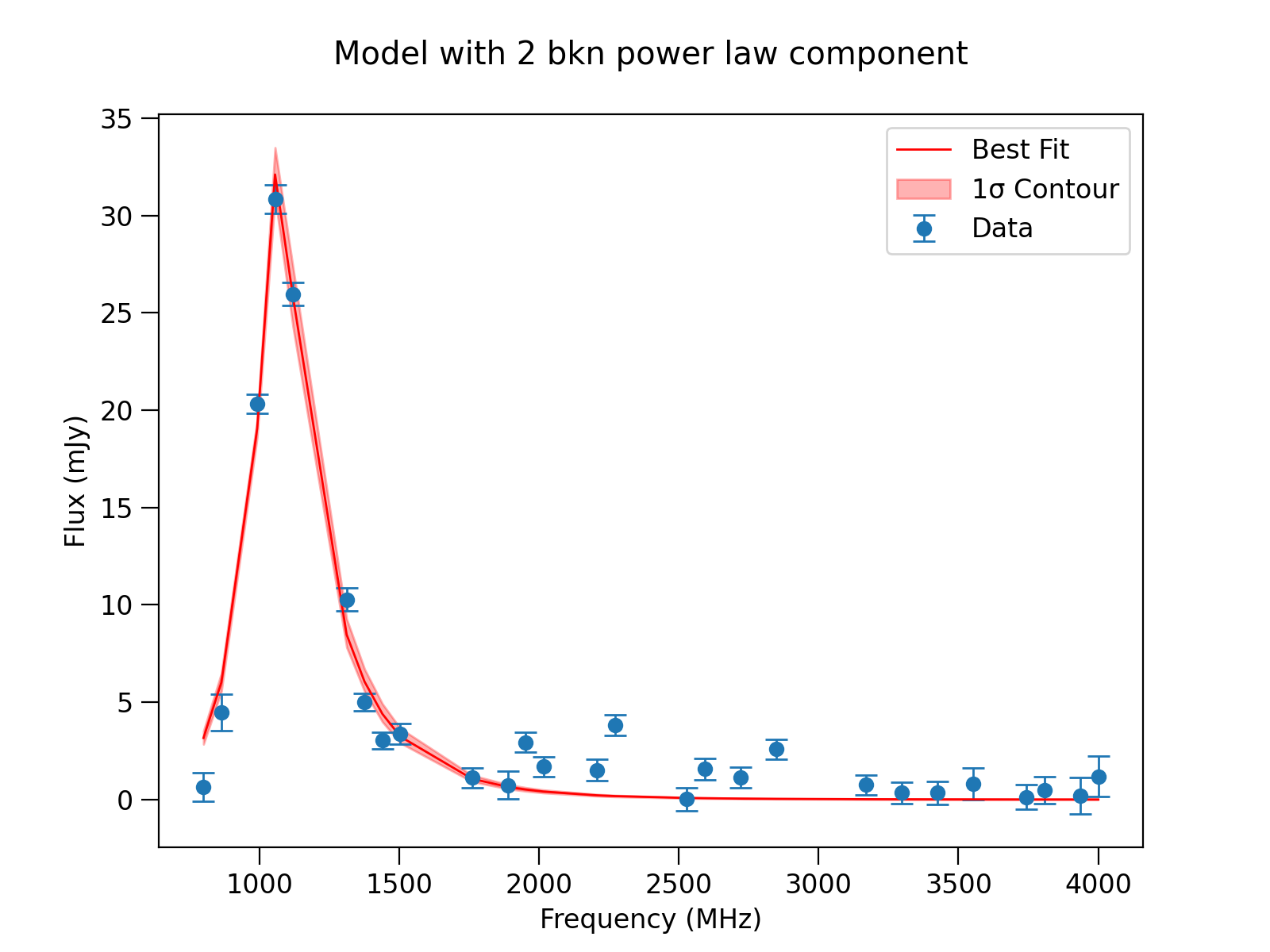}
        \caption{GPS spectral fit for the $2^{nd}$ component.}
        \label{fig:Spectral_fit_plot_interpulse}
    \end{subfigure}
    \hfill
    \begin{subfigure}[t]{0.48\textwidth}
        % \centering
        % \hspace{1.0cm}
        \includegraphics[scale=0.4]{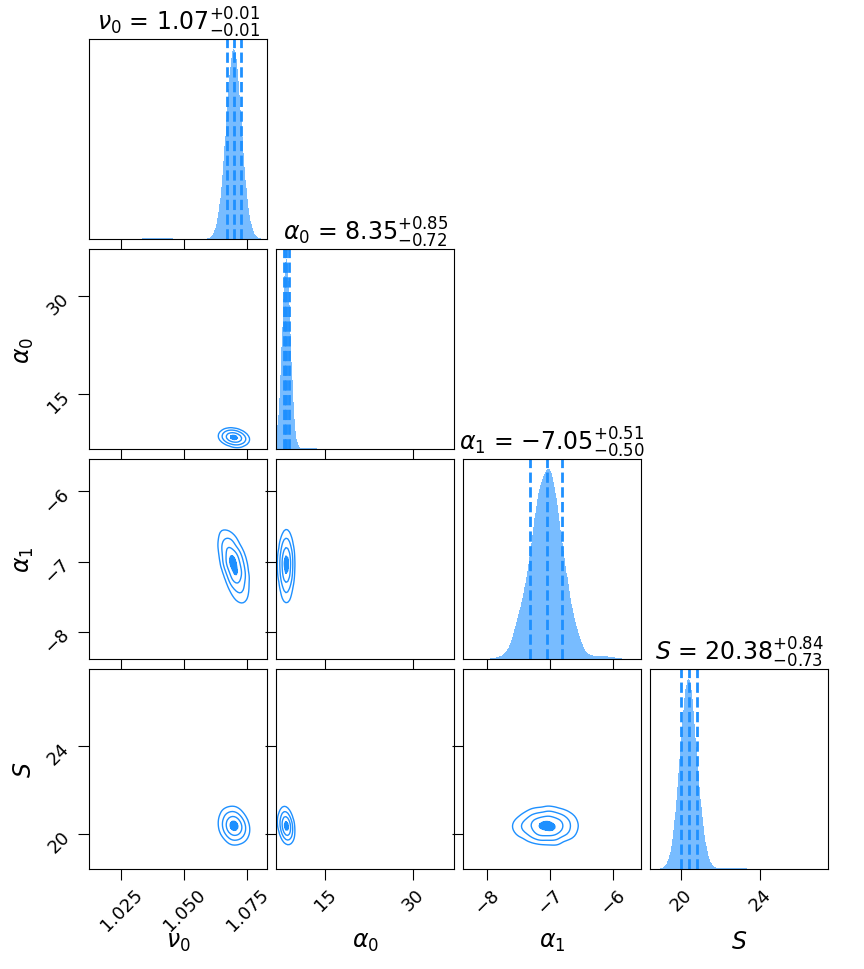}
        \caption{Corner plot for the $2^{nd}$ component.}
        \label{fig:Corner_plot_interpulse}
    \end{subfigure}

    \caption{GPS spectral fits of the power-law model (Equation \ref{eq:Model}) and the corresponding corner plots for the $1^{st}$ and $2^{nd}$ components at orbital phase 0.6 (observing epoch 2023--06--22). Panels (a) and (c) show the best-fit spectra for the $1^{st}$ and $2^{nd}$ components, respectively, while panels (b) and (d) show the corresponding corner plots. The parameter description is same as described in Figure \ref{fig:GPS_eg_corner} or Equation \ref{eq:Model}.\label{fig:GPS_fit_corner_combined}}
    
\end{figure*}

\begin{figure*}[!hbt]
    \begin{subfigure}[t]{0.9\textwidth}
      \centering
      \includegraphics[width=\textwidth]{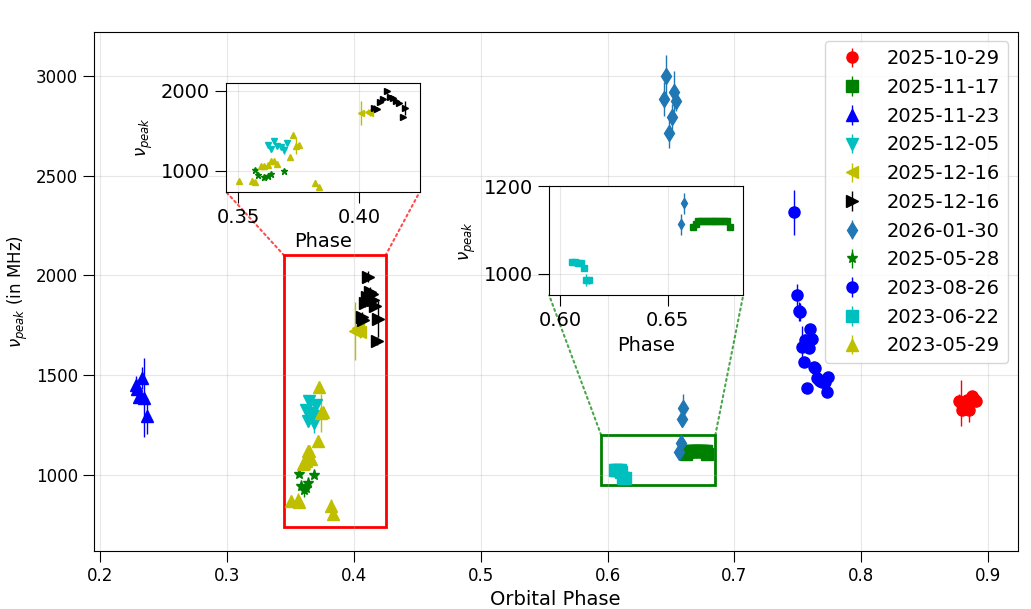}
      \caption{1st component}
      \label{fig:all_epoch_GPS_orbital_phase_main_pulse}
    \end{subfigure}
    \begin{subfigure}[t]{0.9\textwidth}
      \centering
      \includegraphics[width=\textwidth]{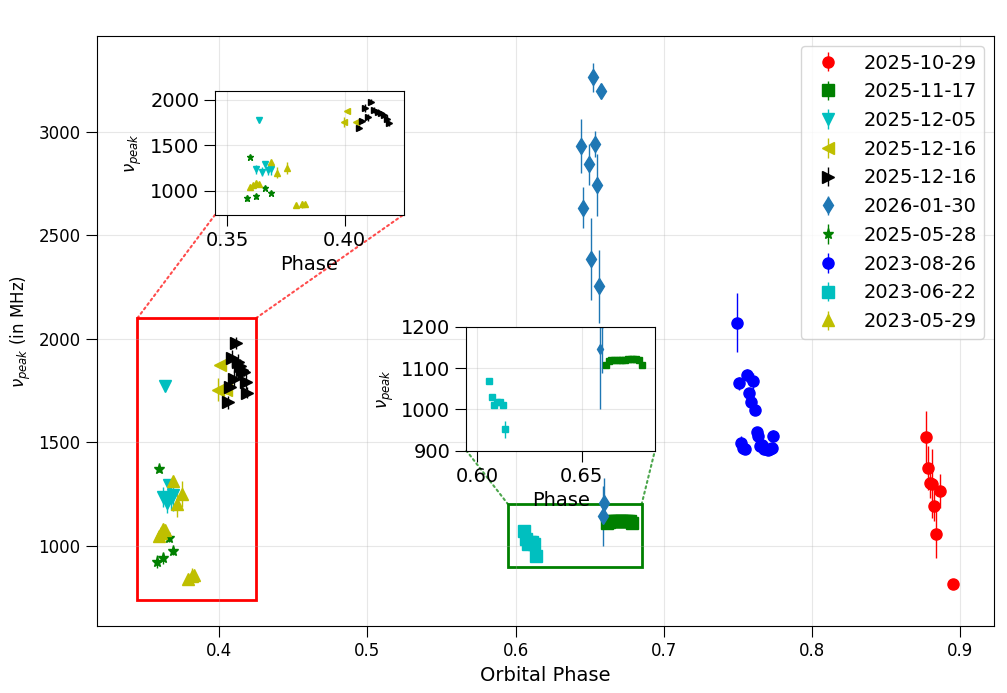}
      \caption{2nd component}
      \label{fig:all_epoch_GPS_orbital_phase_inter_pulse}
    \end{subfigure}
    \caption{
    Variation of the spectral turnover frequency, $\nu_{\rm peak}$, as a function of orbital phase (for the $1^{st}$ component in panel a, and $2^{nd}$ component in panel b). Near orbital phase 0.36 a systematic variation of $\nu_{peak}$ with relatively large scatter is observed. However, near orbital phase 0.6 a relatively smooth trend with lower scatter is observed.\label{fig:all_epoch_GPS_orbital_phase}}
\end{figure*}

The upper panels of Figure~\ref{fig:GPS_fit_corner_combined} present representative Bayesian spectral fits for the $1^{st}$ component, while the lower panels show the corresponding fits for the $2^{nd}$ component obtained from the same 20-minute integration. Repeating this procedure for 20 minutes time-chunks covering all observations listed in Table~\ref{tab:nu_peak_comp_corr} enabled independent measurements of the turnover frequency for both emission components (Figures \ref{fig:all_epoch_GPS_orbital_phase_main_pulse} and \ref{fig:all_epoch_GPS_orbital_phase_inter_pulse}) as a function of orbital phase. %The resulting orbital evolution of $\nu_{peak}$ for the $1^{st}$ and $2^{nd}$ components is presented in Figures~\ref{fig:all_epoch_GPS_orbital_phase_main_pulse} and \ref{fig:all_epoch_GPS_orbital_phase_inter_pulse}, respectively.

Figure~\ref{fig:all_epoch_GPS_orbital_phase_main_pulse} presents the orbital evolution of the spectral turnover frequency, $\nu_{\rm peak}$, for the $1^{st}$ component, obtained by combining measurements from all observing epochs. The turnover frequency exhibits a clear and systematic dependence on orbital phase, increasing from approximately 800 MHz near orbital phase $\sim0.35$, close to the lower edge of the observing band, to nearly 3.0 GHz around orbital phase $\sim0.64$. This substantial evolution demonstrates that the spectral properties of PSR J2144$-$5237 vary continuously as the pulsar traverses its orbit.

%   Specific features of $1^{st}$ component nu_peak versus orbital phase plot

The evolution of $\nu_{\rm peak}$ is remarkably smooth across majority of the sampled orbital phases. Around orbital phases $\sim0.40$ and $\sim0.60$, measurements from independent observing epochs form well-defined sequences with very little scatter, indicating excellent reproducibility of the spectral evolution. A somewhat larger dispersion is observed near orbital phase $\sim0.36$, where measurements cluster around $\nu_{\rm peak}\approx1.3$ GHz. This increased scatter is primarily associated with observations having comparatively lower signal-to-noise ratios. Nevertheless, even at orbital phases sampled during multiple epochs, the turnover frequencies consistently follow the same evolutionary trend, demonstrating that the observed modulation is intrinsic to the binary system rather than arising from observational uncertainties or epoch-to-epoch variability.

%   Desciption of interpulse being similar to main pulse, and the lacuna for encapsulating low SNR chunks

The orbital evolution of the turnover frequency for the $2^{nd}$ component is shown in Figure~\ref{fig:all_epoch_GPS_orbital_phase_inter_pulse}. Owing to the lower flux density of the $2^{nd}$ component, reliable estimates of $\nu_{\rm peak}$ could not be obtained for every 20-minute integration or observing epoch. However, for those orbital phases where robust spectral fits could be obtained, the $2^{nd}$ component exhibits an evolutionary pattern that closely mirrors that of the $1^{st}$ component. The turnover frequency increases systematically with orbital phase over nearly the same frequency range, indicating that both pulse components undergo similar spectral evolution despite their differing intensities.

%   Correlation between interpulse and main pulse nu_peak: significance

To quantify the relationship between the two emission components, we computed the Pearson correlation coefficient between the $\nu_{\rm peak}$ measurements of the $1^{st}$ and $2^{nd}$ components using only those 20-minute integrations with sufficiently high signal-to-noise ratios to ensure reliable parameter estimation. Data segments contaminated by radio-frequency interference (RFI), which resulted in poorly constrained $2^{nd}$ component fits, were excluded from the analysis. The resulting correlation coefficients and associated $p$-values are summarized in Table~\ref{tab:nu_peak_comp_corr}. The strong positive correlations, together with the statistically significant $p$-values, demonstrate that the turnover frequencies of the two pulse components evolve in a highly correlated manner throughout the orbit.

%   Show some of the difference and ratio in main and interpulse nu_peak vs orbital phase for some of the chunks, some explanation (requirement of complex model) on that.

\begin{figure*}[t]
      \centering
      \includegraphics[trim={0 0 0 1.0cm}, clip, scale=0.75]{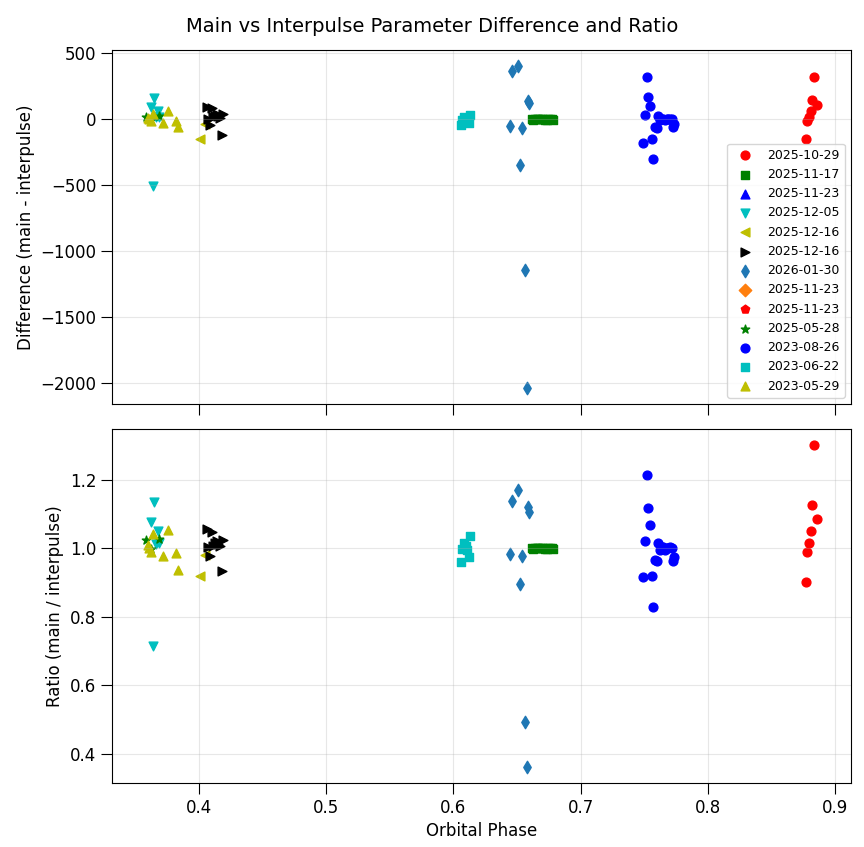}
      \caption{Figure shows the difference and ratio of $\nu_{peak}$ between $1^{st}$ and the $2^{nd}$ pulse components with orbital phase. \label{fig:all_epoch_GPS_orbital_phase_main_IP_diff_ratio}}
      
\end{figure*}

To further examine this correspondence, Figure~\ref{fig:all_epoch_GPS_orbital_phase_main_IP_diff_ratio} presents the difference (upper panel) and the ratio (lower panel) of the turnover frequencies measured for the $1^{st}$ and $2^{nd}$ components. For the majority of the observations, both quantities remain nearly constant throughout the orbit, further supporting the conclusion that the two pulse components experience nearly identical spectral evolution. The largest deviations are observed during the 2026-01-30 epoch and in one 20-minute integration from the 2025-12-05 observation. As seen in Figure~\ref{fig:all_epoch_GPS_orbital_phase_main_IP_diff_ratio}, the $2^{nd}$ component measurements during these intervals exhibit substantially larger scatter, whereas the corresponding main-pulse turnover frequencies remain tightly clustered around $\sim2.7$ GHz. These discrepancies are most likely attributable to limitations of the phenomenological broken power-law model in describing spectra with increased complexity at certain orbital phases. Physically motivated radiative transfer models incorporating frequency-dependent absorption within the magnetized intra-binary plasma are expected to provide a more accurate representation of the observed spectra and further reduce these residual differences.

%   Comment on the similar main and interpulse nu_peal => natural cause to be intra-binary material instead of intrinsic

One of the most striking results of this study is the gradual and continuous evolution of the GPS turnover frequency over a large fraction of the binary orbit. Rather than exhibiting abrupt changes confined to particular orbital phases, $\nu_{\rm peak}$ evolves smoothly from approximately 0.8 GHz to nearly 3.0 GHz and subsequently decreases as the pulsar progresses through its orbit. Measurements obtained during different observing epochs consistently trace the same evolutionary path, indicating that the observed modulation is highly reproducible. This behaviour strongly suggests that the physical conditions along the line of sight evolve continuously with orbital phase, reflecting gradual changes in the density, magnetic field strength, and effective path length through the absorbing plasma. Such smooth spectral evolution is naturally expected if the radio emission propagates through a structured and spatially varying intra-binary medium, whose properties are regulated by the orbital motion of the binary system.

%   Comment on main and interpulse nu_peak similarity from the posibility of intrisic variability. 

The close correspondence between the $1^{st}$ and $2^{nd}$ components provides an additional and important observational constraint. Since the two pulse components are believed to originate from different magnetic poles of the pulsar and therefore represent distinct magnetospheric emission regions, their nearly identical evolution cannot be readily explained by intrinsic changes in the pulsar emission mechanism. Instead, the strong correlation between their turnover frequencies indicates that both components are modified by the same external propagation process. These observations therefore provide compelling evidence that the GHz-peaked spectrum of PSR J2144$-$5237 is predominantly governed by propagation through the magnetized intra-binary environment. The results presented here establish, for the first time, that the GPS turnover frequency can serve as a sensitive diagnostic of the evolving plasma conditions in compact binary systems, opening a new avenue for probing pulsar wind-companion interactions through broadband radio spectroscopy.

\section{Discussion} \label{sec:Discussion}

The continuous orbital evolution of the GPS turnover frequency observed in PSR J2144$-$5237 provides important constraints on the physical mechanism responsible for shaping its radio spectrum. Environmental absorption has long been proposed as the origin of the GHz-peaked spectrum (GPS) phenomenon, with free-free absorption, synchrotron self-absorption, and cyclotron resonance being the principal candidate mechanisms \citep{Lewandowski_et_Al_2015, Kijak_et_al_2011b}. The systematic increase and decrease of $\nu_{peak}$ over the binary orbit indicates that the absorbing medium evolves smoothly with orbital phase, suggesting that the observed spectral turnover is governed by the changing line of sight through the intra-binary plasma.

\subsection{Free-free absorption}
Within the free-free absorption framework, the turnover frequency depends on the optical depth of the absorbing plasma, which is determined primarily by the electron temperature ($T_e$) and the emission measure ($EM$) \citep{Lewandowski_et_Al_2015}. Assuming that the observed spectral turnover is produced by thermal free-free absorption, the measured values of $\nu_{peak}$ can be used to estimate these physical parameters at different orbital phases.

Applying this interpretation to PSR J2144$-$5237 suggests that the inferred values of $T_e$ and $EM$ vary systematically with orbital phase, with similar values recurring at corresponding orbital phases in different observing epochs. Such behaviour is qualitatively consistent with an absorbing medium whose density distribution is regulated by the orbital geometry. However, this interpretation encounters an important difficulty. The broken power-law fits require spectral indices that are considerably steeper than those expected from a simple free-free or synchrotron self-absorption model. Moreover, the gradual migration of the turnover frequency across nearly the entire UWL observing band would require substantial variations in the optical depth over relatively short orbital intervals, making the free-free interpretation less compelling.

These limitations suggest that while thermal absorption may contribute to the observed spectra, it is unlikely to be the dominant mechanism responsible for the pronounced orbital evolution of the GPS turnover frequency.

\subsection{Cyclotron resonance}
An alternative explanation is cyclotron resonance interaction between the pulsar radio emission and magnetized plasma surrounding the companion \citep{Kijak_et_al_2011b}. Unlike free-free absorption, cyclotron resonance naturally predicts a turnover frequency that depends directly on the local magnetic field strength, the Lorentz factor of the absorbing electrons, and the viewing geometry. The resonance frequency is given by,

The $\nu_{peak}$ formulae for cyclotron resonance is as follows:
\begin{equation}
    \nu_{peak} = \frac{l\nu_g}{1 - \frac{v_{||}}{c}cos\theta}
\end{equation}
where $l$ is the harmonics of the $\nu_g$ is the relativistic cyclotron frequency 

\begin{equation}
   \nu_g =  \frac{eB}{2\pi \gamma m_e} = 28 \gamma^{-1}\frac{B}{T} \; GHz
\end{equation} 

is the relativistic cyclotron frequency, $B$ is the magnetic field strength, $\gamma$ is the Lorentz factor of the absorbing electrons, $v_\parallel$ is the component of the particle velocity parallel to the magnetic field, and $\theta$ is the angle between the line of sight and the magnetic field. The denominator represents the relativistic Doppler shift between the plasma rest frame and the observer.

The nature of the cyclotron interaction depends on the electron velocity distribution function \citep{Tsurutani_n_Lakhina_1997, Melrose_1968}. For thermal or Maxwellian distributions, where the velocity derivative is negative, energy is transferred from the electromagnetic wave to the plasma, producing cyclotron absorption. Conversely, non-thermal distributions with a positive velocity derivative transfer energy from the plasma to the radiation field, resulting in stimulated cyclotron emission. Thus, the electron distribution determines whether the resonance appears observationally as absorption or emission.

The systematic orbital evolution of $\nu_{peak}$ observed in PSR J2144$-$5237 strongly favours the cyclotron resonance scenario. As the pulsar moves through its orbit, the line of sight samples different regions of the companion's magnetosphere and shocked pulsar wind, where the magnetic field strength, plasma density, and viewing geometry vary continuously. These gradual changes naturally produce the smooth orbital evolution of the resonance frequency observed in both the $1^{st}$ and $2^{nd}$ component. Furthermore, the nearly identical evolution of $\nu_{peak}$ in both components strongly suggests that the modulation originates in the external propagation medium rather than within the pulsar magnetosphere itself.

The observed evolution of the turnover frequency therefore provides a new diagnostic of the magnetized intra-binary plasma. By modelling the orbital dependence of $\nu_{peak}$ within the cyclotron resonance framework, it is possible to constrain the magnetic field strength of the companion, the Lorentz factor and spatial distribution of the pulsar wind electrons, and the geometry of the absorbing region. Since our observations span $\sim2$ years for a binary system of 10-day orbital period and are unevenly sampled in orbital phase, long-term spectral variations could potentially be attributed to large-scale clumps in the interstellar medium (ISM). However, the observed orbital-phase dependence of parameters such as $\nu_{peak}$, particularly at overlapping orbital phases sampled across multiple epochs, cannot be readily explained by such irregularities in the ISM. Such constraints offer a unique probe of pulsar wind-companion interactions that is complementary to traditional eclipse, polarization, and rotation-measure studies.
% \textbf{In the least likely case of significant spectral variations between epochs caused by clumpy interstellar material, these constraints would still provide a unique probe of pulsar wind–companion interactions, complementary to traditional studies based on eclipses, polarization, and rotation measures.}
%Such constraints offer a unique probe of pulsar wind-companion interactions that is complementary to traditional eclipse, polarization, and rotation-measure studies.

%   Correlation of Main and inter-pulse : High correlation shows extrinsic (rejects intrinsic effects)

\section{Future scope} \label{sec:Future_work}
The present work provides the first systematic evidence for the continuous orbital evolution of the GHz-peaked spectrum (GPS) turnover frequency in a binary millisecond pulsar. While these results strongly support an environmental origin for the observed spectral evolution, they also motivate several avenues for future investigation that can substantially improve our understanding of pulsar wind-companion interactions and the structure of the intra-binary medium.

\begin{itemize}

\item \textbf{Phase-resolved spectral modelling of individual pulse components:} A more detailed investigation of the spectra of individual profile components using full-Stokes observations will allow the spectral behaviour of each emission region to be characterized independently.

\item \textbf{Development of physically motivated spectral models:} The broken power-law model employed in this work provides a convenient phenomenological description of the spectral turnover but does not explicitly account for the underlying absorption physics. Simultaneous fitting of physically motivated models, such as : free-free absorption cyclotron resonance, synchrotron absorption, to orbital-phase-resolved spectra would enable direct estimates of the electron density, plasma temperature, magnetic field strength, and particle energy distribution within the absorbing/emission medium, thereby providing a quantitative description of the binary environment.

\item \textbf{Joint analysis of spectral and polarization evolution:} A combined analysis of the orbital evolution of the GPS turnover frequency with variations in polarization fraction, polarization position angle, rotation measure (RM), and dispersion measure (DM) will enable self-consistent modelling of the binary geometry and magnetic field structure. Such an approach can identify the dominant absorption mechanism while simultaneously constraining the orientation of the companion's magnetic field and the spatial distribution of the absorbing plasma.

\item \textbf{Multi-parameter studies of the intra-binary medium:} Future investigations should combine orbital-phase-resolved measurements of the spectral turnover frequency, eclipse duration, pulse broadening, scattering, DM, RM, flux density, and pulse morphology. Studying the correlations among these observables will provide a comprehensive picture of the physical conditions within the intra-binary medium, including its density distribution, magnetic field topology, turbulence, and temporal variability.

\item \textbf{Application to a larger sample of binary millisecond pulsars:} Extending this analysis to a larger sample of spider pulsars, eclipsing MSPs, and detached binaries using wideband receivers such as the Parkes UWL, MeerKAT, and eventually the Square Kilometre Array (SKA) will establish whether orbital modulation of the turnover frequency is a ubiquitous property of binary MSPs. 

\item \textbf{Three-dimensional modelling of pulsar wind--companion interactions:} Coupling orbital spectral evolution with magnetohydrodynamic (MHD) simulations will reconstruct the density and magnetic-field structure of the shocked intra-binary plasma, linking the observed turnover evolution to pulsar wind--companion interactions and particle acceleration.

\end{itemize}

% \begin{figure*}[t]{0.9\textwidth}
%       \centering
%       \includegraphics[width=\textwidth]{all_epoch_GPS_orbital_phase_main_IP_diff_ratio.png}
%       \caption{Figure shows the difference and ratio of $\nu_{peak}$ between main pulse and inter pulse with orbital phase}
%       \label{fig:all_epoch_GPS_orbital_phase_main_pulse}
% \end{figure*}

\begin{acknowledgments}
We thank the referee for comments that helped to improve
the paper. We thank the Department of Atomic Energy, Government of India, under project No.12-R\&D-TFR-5.02-0700. The data presented in this paper were obtained by the Parkes radio telescope. Murriyang, CSIRO’s Parkes radio telescope, is part of the Australia Telescope National Facility (https://ror.org/05qajvd42) which is funded by the Australian Government for operation as a National Facility managed by CSIRO. We acknowledge the Wiradjuri people as the Traditional Owners of the Observatory site.\end{acknowledgments}

\software{astropy \citep{astropy:2013, astropy:2018, astropy:2022},  
          Dynesty \citep{Dynesty_2020},
          psrqpy \citep{psrqpy}}

\bibliography{sample701}{}
\bibliographystyle{aasjournalv7}

\end{document}